\documentclass[preprint,aps]{revtex4}
\usepackage{amssymb,amsmath,doublespace,boxedminipage,epsfig,graphicx}

\begin{document}

\title{Scaling Properties of Saddle-Node Bifurcations on Fractal Basin 
Boundaries}

\author{Romulus Breban}
\altaffiliation{Department of Physics}
\altaffiliation{Institute for Research in Electronics and Applied Physics.}
\author{Helena E. Nusse} 
\altaffiliation{Institute for Physical Sciences and Technology; 
Permanent Address: University 
of Groningen, Department of Econometrics, P.O.~Box 800, NL-9700 AV, 
Groningen, The Netherlands.}
\author{Edward Ott$^{*,}$}
\altaffiliation{Also at Department of Electrical and Computer Engineering.}
\affiliation{University of Maryland, College Park, Maryland 20742}

\date{\today} 
\begin{abstract}
We analyze situations where a saddle-node bifurcation occurs on a
fractal basin boundary. Specifically, we are interested in what happens when a
system parameter is slowly swept in time through the bifurcation.
Such situations are known to be indeterminate in the sense that it is 
difficult to predict the eventual fate of an orbit that tracks the 
pre-bifurcation node attractor as the system parameter is swept through the 
bifurcation. In this paper we investigate the scaling of (1) the fractal
basin boundary of the static (i.e., unswept) system near the saddle-node 
bifurcation, (2) the dependence of the orbit's final destination on the
sweeping rate, (3) the dependence of the time it takes for an attractor to 
capture a swept orbit on the sweeping rate, and (4) the dependence of the 
final attractor capture probability on the noise level. 
With respect to noise, our main result is that the effect of noise scales
with the 5/6 power of the parameter drift rate.
Our approach is to first investigate all
these issues using one-dimensional map models. The simplification of treatment
 inherent in one dimension
greatly facilitates analysis and numerical experiment, aiding us in obtaining
the new results listed above. Following our one-dimensional investigations,
we explain that these results can be applied to two-dimensional systems. We
show, through numerical experiments on a periodically forced second
order differential equation example, that the scalings we have found
also apply to systems that result in two dimensional maps.

\end{abstract}

\maketitle

\section{Introduction}
It is common for dynamical systems to have two or more
coexisting attractors. In predicting the long-term behavior of a such a system,
it is important to determine sets of initial conditions of
orbits that approach each attractor (i.e., the basins of attraction). The
boundaries of such sets are often fractal (\cite{McDonald}, Chapter 5 of
\cite{Ott}, and references therein).
The fine-scale fractal structure of such a boundary implies
increased sensitivity to errors in the initial conditions: Even a 
considerable decrease in the uncertainty of initial conditions may yield only
 a relatively small decrease in the probability of making an error in
 determining in which basin such an initial condition belongs 
\cite{McDonald,Ott}. For discussion of fractal basin boundaries
in experiments, see Chapter 14 of \cite{Virgin}.

Thompson and Soliman \cite{Thompson1} showed that another
source of uncertainty induced by fractal basin boundaries may arise in
situations in which there is slow (adiabatic) variation of the system. 
For example,
consider a fixed point attractor of a map (a node). As a system parameter
varies slowly, an orbit initially placed on the node attractor moves with
time, closely following the location of the solution for the fixed point in the
absence of the temporal parameter variation. As the parameter varies, the
node attractor may suffer a saddle-node bifurcation. For definiteness, say
that the node attractor exists for values of the parameter $\mu$ in the range
 $\mu<\mu_*$, and that the saddle-node bifurcation of the node occurs at
 $\mu=\mu_*$. Now assume that, for a parameter interval $[\mu_L,\mu_R]$
with $\mu_L <\mu_*< \mu_R$, in addition to the node,
 there are also two other attractors A and B, and that the boundary of
the basin of attractor A, attractor B and the node is a fractal basin 
boundary. We are interested in the typical case where, before 
the bifurcation, the saddle lies on the fractal basin
boundary, and thus, at the bifurcation, the merged saddle-node orbit is on the
 basin boundary. In such a case an arbitrarily
small ball about the saddle-node at $\mu=\mu_*$ contains pieces of the
basins of both A and B. Thus, as $\mu$ slowly increases through $\mu_*$, it is
unclear whether the orbit following the node will go to A or to B after the
node attractor is destroyed by the bifurcation. In practice, noise or 
round-off error may lead the orbit
to go to one attractor or the other, and the result can often depend very
sensitively on the specific value of the slow rate at which the system 
parameter varies.

We note that the study of orbits swept through an indeterminate
saddle-node bifurcation belongs to the theory of dynamical bifurcations.
Many authors have analyzed orbits swept through other bifurcations, like the
period doubling bifurcation \cite{pd}, the pitchfork bifurcation \cite{pf,tc},
and the transcritical
bifurcation \cite{tc}. In all these studies of the bifurcations listed above, 
the local
structure before {\it and} after the bifurcation includes stable invariant
manifolds varying smoothly with the bifurcation parameter (i.e., a stable
fixed point that exists before or after the bifurcation, and whose
location varies smoothly with the bifurcation parameter). This particular
feature of the local bifurcation structure, not shared by the saddle-node
bifurcation, allows for well-posed, locally defined, problems of dynamical
bifurcations. The static saddle-node bifurcation has received much attention
in theory and experiments \cite{Whitley,Pomeau,sn}, but so far, no dynamical 
bifurcation
problems have been defined for the saddle-node bifurcation. In this work,
we demonstrate that, in certain common situations, global structure (i.e., an
invariant Cantor set or a fractal basin boundary) adds to the local properties
of the saddle-node bifurcation and allows for well-posed problems of
dynamical bifurcations.

Situations where a saddle-node bifurcation occurs on a fractal basin
boundary have been studied in two dimensional Poincar\'e maps of damped
forced oscillators \cite{Thompson1,Nusse,Breban}. Several 
examples of such systems are known \cite{Thompson1,Breban}, and
it seems that this is a common occurence in dynamical systems. In
 this work, we first focus on saddle-node bifurcations that occur for one 
parameter families of smooth
one dimensional maps having multiple critical points (a critical point is a
point at which the derivative of the map vanishes). Since one dimensional
dynamics
is simpler than two dimensional dynamics, indeterminate bifurcations can
be more simply studied, without the distraction of extra mathematical
structure. Taking advantage of this, we are able to efficiently investigate
several scaling properties of these 
bifurcations. In particular, we investigate the scaling of (1) the fractal
basin boundary of the static (i.e., unswept) system near the saddle-node 
bifurcation (Secs.~\ref{sec:dim} and \ref{sec:scdim}), (2) the dependence of 
the orbit final destination on the sweeping rate
(Sec.~\ref{sec:sweep}), (3) the dependence of the time it takes
for an attractor to capture the swept orbit following the bifurcation on the
sweeping rate
(Sec.~\ref{sec:capture}), and (4) the dependence of the final attractor 
capture probability on the noise level (Sec.~\ref{sec:noise}). Following our 
one-dimensional investigations, we explain that these results apply to
two dimensional systems.
We show, through numerical experiments on the periodically forced 
Duffing oscillator, that the scalings we have found
also apply to higher dimensional systems (Sec.~\ref{sec:Duff}).

For one-dimensional maps,
a situation dynamically similar to that in which there is indeterminacy in 
which attractor captures the orbit can also occur in cases where there
are two rather than three (or more) attractors (Sec.~\ref{sec:model2}). 
In particular, we can have the situation where one attractor persists for all
values of the parameters we consider, and the other attractor is a node 
which is destroyed via a saddle-node bifurcation on the basin boundary 
separating the 
basins of the two attractors. In such a situation, an orbit starting on the 
node, and swept through the saddle-node bifurcation, will go to the remaining 
attractor. It is possible to distinguish different ways that the orbit 
initially on the node approaches the remaining attractor. We find that the way
in which this attractor is approached can be indeterminate.

\section{Indeterminacy in Which Attractor Is Approached}
\label{sec2}

We consider the general situation of a one dimensional real map $f_\mu(x)$
depending on a parameter $\mu$. We assume the following: (1) the map is twice
differentiable with respect to $x$, and once differentiable with respect to
$\mu$ (the derivatives are continuous); (2) $f_\mu$ has at least two 
attractors sharing a fractal basin boundary for parameter values in the 
vicinity of $\mu_*$; and (3) an attracting fixed point $x_*$ of the map
 $f_\mu(x)$ is destroyed by a saddle-node bifurcation as the parameter $\mu$ 
increases through a critical value $\mu_*$, and this saddle-node bifurcation 
occurs on the common boundary of the basins of the two attractors.

We first recall the saddle-node bifurcation theorem 
(see for example \cite{Whitley}). If the map $f_\mu(x)$ satisfies: (a)
$f_{\mu_*}(x_*)=x_*$, (b) $\frac{\partial f_{\mu_*}}{\partial x}(x_*)=1$,
(c) $\frac{\partial^2 f_{\mu_*}}{\partial^2 x}(x_*)>0$, and (d)
$\frac{\partial f}{\partial \mu}(x_*;\mu_*) >0$, then the map $f_\mu$
undergoes a backward saddle-node bifurcation (i.e., the node attractor
is destroyed at $x_*$ as $\mu$ increases through $\mu_*$). If the 
inequality in either (c) or (d) is reversed, then the map undergoes a 
forward saddle-node bifurcation, while, if both these inequalities are 
reversed, the bifurcation remains backward. A saddle-node bifurcation in a 
one dimensional map is also called a tangent or a fold bifurcation.

\subsection{Model}
\label{model1}

As an illustration of an indeterminate saddle-node bifurcation in a 
one-dimensional map, we construct an example in the following way.
We consider the logistic map for a parameter value where there is a stable
period three orbit. We denote this map $g(x)$ and its third iterate 
$g^{[3]}(x)$. The map $g^{[3]}(x)$ has three stable fixed points. We 
perturb the map $g^{[3]}(x)$ by adding a function (which depends on a
parameter $\mu$)
that will cause a saddle-node bifurcation of one of the attracting fixed
points but not of the other two [see Figs.~1(a) and 1(b)]. We investigate
\begin{eqnarray}
\label{eq:model1}
f_\mu(x)=g^{[3]}(x)+\mu\sin(3\pi x),\quad \mbox{where}
\quad g(x)=3.832\,x(1-x).
\end{eqnarray}
Numerical calculations show that the function $f_\mu(x)$ satisfies all
the conditions of the saddle-node bifurcation theorem for having a backward
saddle-node bifurcation at $x_*\approx 0.15970$ and $\mu_*\approx 0.00279$.
Figure 2(a) displays how the basins of the three attracting fixed 
points of the map $f_\mu$ change with variation of $\mu$. For
$\mu=0$ the third iterate of the logistic map is unperturbed, and it has three
attracting fixed points whose basins we color-coded with blue, green and red.
For every value of $\mu$, the red region $R[\mu]$ is the set of initial
conditions attracted to the rightmost stable fixed point which we denote 
$R_\mu$. The green region
$G[\mu]$ is the set of initial conditions attracted to the middle stable
fixed point which we denote $G_\mu$. The 
blue region $B[\mu]$ is the set of initial conditions
attracted to the leftmost stable fixed point which we denote $B_\mu$.

For $\mu<\mu_*$, each of these colored sets has infinitely many
disjoint intervals and a fractal boundary.
As $\mu$ increases, the leftmost stable fixed point $B_\mu$ is
destroyed via a saddle-node bifurcation on the fractal basin boundary.
In fact, in this case, for $\mu <\mu_*$, every boundary point of one basin is 
a boundary
point for all three basins. (That is, an arbitrarily small $x$-interval
centered about any point on the boundary of any one of the basins contains
pieces of the other two basins.)
The basins are so-called Wada basins \cite{Wada}.
This phenomenon of a saddle-node bifurcation on the fractal boundary
of Wada basins also occurs for the damped forced oscillators studied in
Refs.~\cite{Nusse,Breban}. Alternatively, if we look at the saddle-node
bifurcation as $\mu$ decreases through the value $\mu_*$,
 then the basin $B[\mu]$ of the newly created stable fixed point
immediately has infinitely many disjoint intervals and its boundary displays
fractal structure. According to the terminology of Robert et 
al.~\cite{Robert}, 
we may consider this bifurcation an example of an `explosion'.

\subsection{Dimension of the Fractal Basin Boundary}
\label{sec:dim}
Figure 3 graphs the computed dimension $D$ of the fractal basin
boundary versus the parameter $\mu$. For $\mu<\mu_*$, we observe that $D$
appears to be a
continuous function of $\mu$. Park et al.~\cite{Park} argue that the fractal
dimension of the basin boundary near $\mu_*$, for $\mu<\mu_*$, scales as
\begin{equation}
D(\mu)\approx D_*-k(\mu_*-\mu)^{1/2},
\end{equation}
with $D_*$ the dimension at $\mu=\mu_*$ ($D_*$ is less than the 
dimension of the phase space), and $k$ a positive constant.
Figure 3 shows that the boundary dimension $D$ experiences a discontinuous
jump at the saddle-node bifurcation when $\mu=\mu_*$.
We believe that this is due to the fact that the basin $B[\mu]$ suddenly
disappears for $\mu >\mu_*$.

The existence of a fractal basin boundary has important practical
consequences. In particular, for the purpose of determining which attractor
eventually captures a given orbit, the arbitrarily fine-scaled structure of
fractal basin boundaries implies considerable sensitivity to small errors in
initial conditions. If we assume that initial points
cannot be located more precisely than some $\epsilon > 0$, then we cannot
determine which basin a point is in, if it is within $\epsilon$  of the basin
boundary. Such points are called $\epsilon$-uncertain. The Lebesgue measure
of the set of $\epsilon$-uncertain points (in a bounded region of
interest) scales like $\epsilon^{D_0-D}$, where $D_0$ is the dimension of
the phase space ($D_0=1$ for one dimensional maps) and $D$ is the box-counting
dimension of the basin boundary \cite{McDonald}. For the case of a fractal
basin boundary $(D_0-D)<1$. When $D_0-D$ is 
small, a large decrease in $\epsilon$ results in a relatively small decrease
 in $\epsilon^{D_0-D}$. This is discussed in Ref.~\cite{McDonald} 
which defines the uncertainty dimension, $D_u$, as follows.
Say we randomly pick an initial condition $x$ with uniform probability
density in a state-space region $S$. Then we randomly pick another initial 
condition $y$ in $S$, such that $|y-x|<\epsilon$. Let $p(\epsilon,S)$ be
 the probability that $x$ and $y$ are in different basins. [We can think of
$p(\epsilon,S)$ as the probability that an error will be made in
determing the basin of an initial condition if the initial
condition has uncertainty of size $\epsilon$.] The uncertainty
dimension of the basin boundary $D_u$ is defined as the limit of
$\ln p(\epsilon,S)/\ln(\epsilon)$ as $\epsilon$ goes to zero \cite{McDonald}.
Thus, the probability of error scales as 
$p(\epsilon,S)\sim\epsilon^{D_0-D_u}$,
where for fractal basin boundaries $D_0-D_u<1$.  This indicates enhanced 
sensitivity to small uncertainty in initial conditions.
For example, if $D_0-D_u=0.2$, then a decrease of the initial condition 
uncertainty $\epsilon$ by a factor of 10 leads to only a relative small 
decrease in the final
state uncertainty $p(\epsilon,S)$, since $p$ decreases by a factor of
about $10^{0.2}\approx 1.6$. Thus, in practical terms, it may be
essentially impossible to significantly reduce the final state
uncertainty. In Ref.~\cite{McDonald} it was
conjectured that the box-counting dimension equals the uncertainty dimension
for basin boundaries in typical dynamical systems. In Ref.~\cite{NY} it is
proven that the box-counting dimension, the uncertainty dimension and the
Hausdorff dimension are all equal for the basin boundaries of one and two
dimensional systems that are uniformly hyperbolic on their basin boundary.

We now explain some aspects of the character of the dependence of $D$ on $\mu$
(see Fig.~3). From Refs.~\cite{PVT} it follows that the box-counting dimension
and the Hausdorff
dimension coincide for all intervals of $\mu$ for which the map $f_\mu$ is
hyperbolic on the basin boundary, and that the dimension depends
continuously on the parameter $\mu$ in these intervals. For $\mu>\mu_*$,
there are many parameter values for which the map has a saddle-node
bifurcation of a periodic orbit on the fractal basin boundary. At such
parameter values, which we refer to as saddle-node bifurcation parameter
values, the dimension is expected to be discontinuous (as it is at the
saddle-node bifurcation of the fixed point, $\mu=\mu_*$, see Fig.~3). 
In fact, 
there exist sequences of saddle-node bifurcation parameter values converging 
to $\mu_*$ \cite{MK}.
Furthermore, for each parameter value $\mu>\mu_*$ for which the map
undergoes a saddle-node bifurcation, there exists a sequence of saddle-node
bifurcation parameter values converging to that parameter value.
The basins of attraction of the periodic orbits created by saddle-node
bifurcations of high period exist only for very small
intervals of the parameter $\mu$. We did not encounter them numerically
by iterating initial conditions for a discrete set of
values of the parameter $\mu$, as we did for the basin of our fixed point
attractor.

\subsection{Scaling of the Fractal Basin Boundary}
\label{sec:scdim}
Just past $\mu_*$, the remaining green and red basins display an alternating
stripe structure [see Fig.~2(b)]. The red and green stripes are interlaced in
a fractal structure. As we approach the
bifurcation point, the interlacing becomes finer and finer scaled, with the
scale approaching zero as $\mu$ approaches $\mu_*$. Similar fine
scaled structure is present in the neighborhood of all preiterates of $x_*$.
 If one changes the horizontal axis of
Figs.~2(a,b) from $\mu$ to $(\mu-\mu_*)^{-1/2}$, then, the complex
alternating stripe structure appears asymptotically periodic
[see Fig.~4(a)].
[Thus, with identical horizontal scale, the dimension plot in Fig.~4(b) 
appears asymptotically
periodic, as well.] We now explain why this is so. We restrict
our discussion to a small neighborhood of $x_*$. Consider
the second order expansion of $f_\mu$ in the vicinity of $x_*$ and $\mu_*$
\begin{equation}
\label{eq:can}
\hat{f}_{\hat\mu}(\hat x)=\hat{\mu}+\hat{x}+a\hat{x}^2, \quad\mbox{where }
\left\{
\begin{array}{l} \hat{x}=x-x_*,\\ \hat{\mu}=\mu-\mu_*, \end{array}\right.
\end{equation}
and $a\approx 89.4315$.
The trajectories of $\hat{f}_{\hat\mu}$ in
the neighborhood of $\hat x=0$, for $\hat\mu$ close to zero, are good 
approximations to
trajectories of $f_\mu$ in the neighborhood of $x=x_*$, for $\mu$ close
to $\mu_*$. Assume that we start with a certain initial condition for
$\hat{f}_{\hat\mu}$, $\hat x_0=\hat x_s$, and we ask the following question:
What are all the positive values of the parameter $\hat\mu$ such
that a trajectory
passes through a fixed position $\hat x_f>0$ at some iterate $n$?
For any given $x_f$ which is not on the fractal basin boundary, there exists 
a range of $\mu$ such that iterates
of $x_f$ under $f_\mu$ evolve to the same final attractor, for all
values of $\mu$ in that range. In particular, once $a\hat x^2$ appreciably
exceeds $\hat\mu$, the subsequent evolution is approximately independent of
$\hat\mu$. Thus, we can choose $\hat x_f\gg \sqrt{\hat\mu/a}$, but still
small enough so that it lies in the region of validity of the canonical
form \eqref{eq:can}. There exists a range of such $\hat x_f$ values satisfying
these requirements provided that $|\hat\mu|$ is small enough.

Since consecutive iterates of $\hat{f}_{\hat\mu}$ in
the neighborhood of $\hat x=0$ for $\hat\mu$ close to zero differ only
slightly, we approximate the one dimensional map,
\begin{equation}
\hat x_{n+1}=\hat f_{\hat\mu}(\hat x_n)=\hat\mu+\hat x_n+a\hat x^2_n,
\end{equation}
by the differential equation \cite{Pomeau},
\begin{equation}
\label{eq:eqdiff}
\frac{d\hat x}{dn}=\hat\mu+a \hat x^2,
\end{equation}
where in \eqref{eq:eqdiff} $n$ is considered as a continuous, rather than
a discrete, variable.
Integrating \eqref{eq:eqdiff} from $\hat x_s$ to $\hat x_f$ yields
\begin{equation}
\label{trans2}
n\sqrt{a\hat\mu}={\rm arctan}
\left(\sqrt{\frac{a}{\hat\mu}}\hat x_f\right)-{\rm arctan}
\left(\sqrt{\frac{a}{\hat\mu}}\hat x_s\right).
\end{equation}
Close to the saddle-node bifurcation (i.e., $0<\hat\mu\ll 1$, and
$\hat x_{s,f}$ close to zero), $\hat{f}_{\hat\mu}$ is a good approximation to
$f_\mu$. For $|\hat x_{s,f}|\sqrt{(a/\hat\mu)}\gg 1$ Eq.~\eqref{trans2} 
becomes
 \begin{equation}
\label{trans_}
n\sqrt{a\hat\mu}\approx\pi.
\end{equation}
The values of $\hat{\mu}^{-1/2}_n$ satisfying Eq.~\eqref{trans_} increase
with $n$ in step of $\sqrt{a}/\pi$. For our example
we have $a\approx 89.4315$, thus $\sqrt{a}/\pi\approx 3.010$. Counting many
periods like those in Fig.~4 in the region of $x_c$, the
closest critical point to $x_*$ [see Fig.~4(a)], we
find that the period of the stripe structure is 3.015, which is in good
agreement with our theoretical value.

In order to investigate the structure of the fractal basin boundary
in the vicinity of the saddle-node bifurcation 
(i.e., $\hat x_s$ close to $\hat x_*=0$),
we consider \eqref{trans2} in the case where we demand only
$|\hat x_{f}|\sqrt{(a/\hat\mu)}\gg 1$. Thus, Eq.~\eqref{trans2} becomes
\begin{equation}
\label{trans}
n\sqrt{a\hat\mu}\approx\frac{\pi}{2}-{\rm arctan}
\left(\sqrt{\frac{a}{\hat\mu}}\hat x_s\right).
\end{equation}
Let $\hat{\mu}^{-1/2}_n(\hat{x}_s)$ denote the solution of Eq.~\eqref{trans}
for $\hat \mu$.
Equation \eqref{trans} implies the behavior of
$\hat{\mu}^{-1/2}_n(\hat{x}_s)$ as function of $\hat{x}_s$ and $n$ as
sketched in Fig.~5. For a fixed $n$, $\hat{\mu}^{-1/2}_n$ has
a horizontal asymptote at the value $n\sqrt{a}/\pi$ as
$\hat{x}_s\rightarrow -\infty$, and a vertical asymptote to infinity at
$\hat{x}_s=1/(an)$. For $\hat{x}_s<0$, we have an infinite number of values
of the parameter $\hat{\mu}$, for which an orbit of $\hat{f}_{\hat\mu}$
starting at $\hat{x}_s$ passes through the same position $\hat{x}_f$, after
some number of iterations.  For $\hat{x}_s=0$ (i.e., $x_s=x_*$), we also have
an infinite number of $\hat{\mu}^{-1/2}_n(0)$, but with constant step
$2\sqrt{a}/\pi$ rather than $\sqrt{a}/\pi$
(see the intersections marked with black dots in the Fig.~5). This is
hard to verify from numerics, since $\frac{\partial \hat{\mu}^{-1/2}_n}
{\partial \hat{x}_s}(0)=a^{3/2}(2n/\pi)^2$ increases with $n^2$, and the
stripes become very tilted in the neighborhood of 
$\hat{x}_s=\hat{x}_*=0$. [See Fig.~4(a), 
where the approximate positions of $x_c$ and $x_*$ on the vertical axis
are indicated.]
For $\hat{x}_s > 0$, $\hat{\mu}^{-1/2}_n$ has only a limited
number of values with $n_{\rm max} < 1/(a\hat{x}_0)$.

\subsection{Sweeping Through an Indeterminate Saddle-Node Bifurcation}
\label{sec:sweep}

In order to understand the consequences of a saddle-node bifurcation on
a fractal basin boundary for systems experiencing slow drift, we imagine the
following experiment. We start with the dynamical system $f_\mu$ at
parameter $\mu_s<\mu_*$, with $x_0$ on the attractor to be destroyed at
$\mu=\mu_*$ by a saddle-node bifurcation (i.e., $B_\mu$). Then, as we iterate,
 we slowly change $\mu$ by
a small constant amount $\delta\mu$ per iterate, thus increasing $\mu$
from $\mu_s$ to $\mu_f>\mu_*$,
\begin{eqnarray}
\label{eq:2Dmap_}
x_{n+1}&=&f_{\mu_n}(x_n),\\
\mu_{n}&=&\mu_s+n\,\delta\mu\nonumber.
\end{eqnarray}
When $\mu \geq \mu_f$ we stop sweeping the parameter $\mu$, and, by iterating
 further, we determine to which of the remaining attractors of $f_{\mu_f}$
 the orbit goes. Numerically, we observe that, if $(\mu_f-\mu_*)$ is not too
 small, then, by the time $\mu_f$ is reached, the orbit is close to the
 attractor of $f_{\mu_f}$ to which it goes. [From our subsequent analysis,
 `not too small $|\mu_{s,f}-\mu_*|$'  translates to choices of $\delta\mu$ 
that
satisfy $(\delta\mu)^{2/3}\ll |\mu_{s,f}-\mu_*|$.] We repeat this for
different values of $\delta\mu$ and we graph the final attractor position
for the orbit versus $\delta\mu$ [see Fig.~6(a)]. For
convenience in the graphical representation of Figs.~6(a,b), we have 
represented
the attractor of the green region $G[\mu]$, denoted $G_{\mu_f}$, as a 0, and 
the attractor of the red region $R[\mu]$, denoted $R_{\mu_f}$, as a 1. 
In Fig.~6(a) we use of 25,000
points having the vertical coordinate either 0 or 1, which we 
connect with straight lines. In an interval of $\delta\mu$ for which
the system reaches the same final attractor (either 0 or 1), the
lines connecting the points are horizontal. Such
intervals appear as white bands in Fig. 6, if they are wider than the width
of the plotted lines connecting 0's and 1's. For example, in
Fig 6(a), the white band centered at $\delta\mu=0.8\times 10^{-3}$ has at the 
bottom a thick horizontal line, which indicates that for
the whole of that interval, the orbit reaches the attractor $G_{\mu_f}$ which 
we represented by 0. Adjacent intervals of width less than the plotted
lines appear as black bands. Within such black bands, an uncertainty in
$\delta\mu$ of size equal to the width of the plotted line makes the attractor
that the orbit goes to indeterminate. Figure 6(a) shows that the widths of the
white bands decrease as $\delta\mu$ decreases, such that, for small
$\delta\mu$, we see only black.

If $(\mu_f-\mu_*)$ is large enough (i.e., 
$(\delta\mu)^{2/3}\ll |\mu_{f}-\mu_*|$),
 numerics and our subsequent analysis show that Fig. 6 is independent of 
$\mu_f$. This fact can be understood as follows. Once $\mu=\mu_f$, the orbit 
typically lands in the green or the red basin of attraction and goes to the
corresponding attractor. Due to sweeping, it is possible for the orbit to
switch from being in one basin of attraction of the {\it time-independent}
map $f_\mu$ to the other, since the basin
boundary between $G[\mu]$ and $R[\mu]$ changes with $\mu$. However,
the sweeping of $\mu$ is slow (i.e., $\delta\mu$ is small), and, once
$(\mu-\mu_*)$ is large enough, the orbit is far enough from the fractal
basin boundary, and the fractal basin boundary changes too little to switch
the orbit between $G[\mu]$ and $R[\mu]$.

We also find numerically that Figs.~6(a,b) are independent of the initial
condition $x_0$, provided that it is in the blue basin $B[\mu_s]$,
sufficiently far from the fractal basin boundary, and that $|\mu_s-\mu_*|$ is 
not too small (i.e., $(\delta\mu)^{2/3}\ll |\mu_s-\mu_*|$).

If one changes the horizontal scale of Fig.~6(a) from $\delta\mu$ to
$1/\delta\mu$ [see Fig.~6(b)], the complex band structure appears
asymptotically periodic. Furthermore, we find that the period in
$(1/\delta\mu)$ of the structure in Fig.~6(b) asymptotically approaches
$-1/(\mu_s-\mu_*)$ as $\delta\mu$ becomes small.

In order to explain this result, we again consider the map 
$\hat{f}_{\hat\mu}$, the local approximation of $f_\mu$ in the
region of the saddle-node bifurcation. Equations \eqref{eq:2Dmap_} can be
approximated by
\begin{eqnarray}
\label{eq:2Dmap}
\hat x_{n+1}&=&\hat f_{\hat\mu_n}(\hat x_n)=\hat\mu_n+\hat x_n+a\hat x^2_n,\\
\hat\mu_{n}&=&\hat\mu_s+n\,\delta\mu\nonumber.
\end{eqnarray}
We perform the following numerical experiment. We consider orbits of our
approximate two dimensional map given by Eq.~\eqref{eq:2Dmap} starting at
$\hat x_s=-\sqrt{-\hat\mu_s/a}$. We define a final state function of an
orbit swept with parameter $\delta\mu$ in the following way. It
is 0 if the orbit has at least one iterate in a specified fixed interval far 
from the
saddle-node bifurcation, and is 1, otherwise. In particular, we take the 
final state of a swept orbit to be 0 if there exists 
$n$ such that $100<\hat x_n<250$, and to be 1 otherwise.
Figure 6(c) graphs the corresponding numerical results.
Similar to Fig.~6(b), we observe periodic behavior in $1/\delta\mu$ with
period $-1/\hat\mu_s$. In contrast to Fig.~6(b) where the white band structure
seems fractal, the structure within each period in Fig.~6(c) consists of only 
one interval where the final  state is 0 and one interval where the final 
state is 1. This is because $100<\hat x<250$ is a single interval, while the
green basin [denoted 0 in Fig.~6(b)] has an infinite number of disjoint 
intervals and a fractal boundary (see Fig.~2).\\
 
With the similarity between Figs.~6(b) and 6(c) as a guide, we are now in a
position to give a theoretical analysis explaining the observed periodicity in
 $1/\delta\mu$. In particular, we now know that this can be
explained using the canonical map \eqref{eq:2Dmap}, and that the periodicity
result is thus universal [i.e., independent of the details of our particular
example, Eq.~\eqref{eq:model1}]. For slow sweeping (i.e., $\delta\mu$ small),
consecutive iterates of \eqref{eq:2Dmap} in the vicinity of
$\hat x=0$ and $\hat \mu=0$ differ only slightly, and we further approximate
the system by the following Ricatti differential equation,
\begin{eqnarray}
\label{eq:Ricatti}
\frac{d\hat x}{dn}=\hat\mu_s+n\delta\mu+a\hat x^2.
\end{eqnarray}
The solution of Eq.~\eqref{eq:Ricatti} can be expressed in terms of the Airy
functions $Ai$ and $Bi$ and their derivatives, denoted by $Ai'$ and $Bi'$,
\begin{eqnarray}
\label{eq:Ai}
\hat x(n)=\frac{\eta Ai'(\xi)+Bi'(\xi)}{\eta Ai(\xi)+Bi(\xi)}\,
\left(\frac{\delta\mu}{a^2}\right)^{1/3},
\end{eqnarray}
where
\begin{eqnarray}
\xi(n)=-a^{1/3}\,\frac{\hat\mu_s+n\,\delta\mu}{\delta\mu^{2/3}},
\end{eqnarray}
and $\eta$ is a constant to be determined from the initial condition.
We are only interested in the case of slow sweeping,
$\delta\mu\ll 1$, and $\hat x(0)\equiv\hat x_s=-\sqrt{-\hat\mu_s/a}$
(which is the stable fixed point of $\hat f_{\hat\mu}$ destroyed by the
saddle-node bifurcation at $\hat\mu=0$). In particular, we will consider
the case where $\hat\mu_s<0$ and $|\hat\mu_s|\gg\delta\mu^{2/3}$ 
(i.e., $|\xi(0)|\gg 1$). Using $\hat x(0)=-\sqrt{-\hat\mu_s/a}$ to solve for 
$\eta$ yields $\eta\sim {\cal O}[\xi(0)e^{2\xi(0)}]\gg 1$. For positive large
values of $\xi(n)$ (i.e., for $n$ small enough), using the corresponding
asymptotic expansions of the Airy functions \cite{Abramowitz}, the lowest 
order in $\delta\mu$ approximation to \eqref{eq:Ai} is
\begin{eqnarray}
\label{eq:Ri}
\hat x(n)\approx-\sqrt{-\frac{\hat\mu_s+n\,\delta\mu}{a}},
\end{eqnarray}
with the correction term of higher order in $\delta\mu$ being negative.
Thus, for $n$ sufficiently smaller than $-\hat\mu_s/\delta\mu$, the swept
orbit lags closely behind the fixed point for $\hat f_{\hat\mu}$ with
$\hat\mu$ constant. For $\xi\leq 0$, we use the fact that $\eta$ is large to
approximate \eqref{eq:Ai} as
\begin{eqnarray}
\label{eq:Ric}
\hat x(n)\approx\frac{Ai'(\xi)}{Ai(\xi)}\,
\left(\frac{\delta\mu}{a^2}\right)^{1/3}.
\end{eqnarray}
Note that
\begin{eqnarray}
\label{eq:Ric_}
\hat x(-\hat\mu_s/\delta\mu)\approx\frac{Ai'(0)}{Ai(0)}\,
\left(\frac{\delta\mu}{a^2}\right)^{1/3}=
(-0.7290...)\left(\frac{\delta\mu}{a^2}\right)^{1/3}
\end{eqnarray}
gives the lag of the swept orbit relative to the fixed point attractor
evaluated at the saddle-node bifurcation. Equation \eqref{eq:Ric} does not 
apply for $n>n_{\rm max}$, where $n_{\rm max}$ is the value of $n$ for which
$\xi(n_{\rm max})=\tilde\xi$, the largest root of $Ai(\tilde\xi)=0$ (i.e.,
$\tilde\xi=-2.3381...$). At $n=n_{\rm max}$, the normal form approximation
predicts that the orbit diverges to $+\infty$. Thus, for $n$ near 
$n_{\rm max}$, the
normal form approximation of the dynamical system ceases to be valid. Note,
however, that \eqref{eq:Ric} can be valid even for $\xi(n)$ close to
$\xi(n_{\rm max})$. This is possible because $\delta\mu$ is small.
In particular, we can consider times up to the time $n'$ where $n'$ is 
determined
by $\xi'\equiv\xi(n')=\tilde\xi+\delta\xi$, ($\delta\xi>0$ is small,)
provided $|\hat x(n')|\ll 1$ so that the normal form applies. That is, we 
require
$[Ai'(\xi')/Ai(\xi')](\delta\mu/a^2)^{1/3}\ll 1$, which can be satisfied
even if $[Ai'(\xi')/Ai(\xi')]$ is large. Furthermore, we will take the
small quantity $\delta\xi$ to be not too small (i.e.,
$\delta\xi/(a\,\delta\mu)^{1/3}\gg 1$), so that $(n_{\rm max}-n')\gg 1$.
We then consider \eqref{eq:Ric} in the range,
$-(\hat\mu_s/\delta\mu)\leq n<n'$, where the normal form is still valid.

We use Eq.~\eqref{eq:Ric} for answering the following question: What are
all the values of the parameter $\delta\mu$ ($\delta\mu$ small) for which an
orbit passes exactly through the same position $\hat x_f>0$, at some iterate
$n_f$? All such orbits would further evolve to the same final
attractor, independent of $\delta\mu$, provided  
$a\hat x_f^2\gg \hat\mu_s+n_f\,\delta\mu$; i.e., $\hat x_f$ is large enough
that $\hat\mu_f=\hat\mu_s+n_f\,\delta\mu$ does not much influence the orbit 
after $\hat x$ reaches $\hat x_f$. 
[Denote $\xi(n_f)$ as $\xi(n_f)\equiv \xi_f$.] Using \eqref{eq:Ric} we can
estimate when this occurs, $a\hat x_f^2=[Ai'(\xi_f)/Ai(\xi_f)]^2
(\delta\mu^2/a)^{1/3}\gg (\hat\mu_s+ n_f\,\delta\mu)$ or
$[Ai'(\xi_f)/Ai(\xi_f)]^2\gg \xi_f$. This inequality is satisfied when
$\xi_f$ gets near $\tilde\xi$, which is the largest zero of $Ai$
(i.e., $\xi_f=\tilde\xi+\delta\xi$, where $\delta\xi$ is a small positive 
quantity). We now rewrite Eq.~\eqref{eq:Ric} in the following way
\begin{eqnarray}
\label{eq:Ric2}
\frac{1}{\delta\mu}=-\frac{n_f}{\hat\mu_s-
\left[\frac{(\delta\mu)^2}{a}\right]^{1/3}K\left[
\left(\frac{a^2}{\delta\mu}\right)^{1/3}\,\hat x_f\right]},
\end{eqnarray}
representing a transcedental equation in $\delta\mu$ where $\hat \mu_s$ and
$\hat x_f$ are fixed, $n_f$ is a large positive integer
(i.e., $n_f-1$ is the integer part of $(\hat \mu_f-\hat \mu_s)/\delta\mu$), 
and $K(\zeta)$ is the inverse function of $Ai'(\xi)/Ai(\xi)$
 in the neighborhood of $\zeta=(a^2/\delta\mu)^{1/3}\,\hat x_f\gg 1$. Thus
$|K[(a^2/\delta\mu)^{1/3}\,\hat x_f]|\lesssim |K(\infty)|=|\tilde\xi|$.
The difference $[1/\delta\mu(x_f,n_f+1)-1/\delta\mu(x_f,n_f)]$, where 
$\delta\mu(x_f,n_f)$ is the solution of Eq.~\eqref{eq:Ric2},
yields the limit period of the attracting state versus $1/\delta\mu$ graph
(see Fig.~6). We denote this limit period by $\Delta\,(1/\delta\mu)$. For 
small $\delta\mu$,
the term involving $K[(a^2/\delta\mu)^{1/3}\,\hat x_f]$ in Eq.~\eqref{eq:Ric2}
can be neglected, and we get 
$\Delta\,(1/\delta\mu)=-\hat\mu_s^{-1}=(-\mu_s+\mu_*)^{-1}$.
Figure 7 graphs numerical results of $[\Delta\,(1/\delta\mu)]^{-1}$ versus
$\mu_s$ for our map example given by Eq.~\eqref{eq:2Dmap_}. The fit line is 
$[\Delta\,(1/\delta\mu)]^{-1}=-0.9986\mu_s+0.0028$,
which agrees well with the prediction of the above analysis and our
numerical value for $\mu$ at the bifurcation, $\mu_*\approx 0.00279$.

An alternate point of view on this scaling property is as follows.
For $\hat\mu <0$ (i.e., $\mu<\mu_*$)
and slow sweeping (i.e., $\delta\mu$ small), the orbit closely follows the
stable fixed point attractor of $\hat{f}_{\hat\mu}$, until $\hat\mu \geq 0$,
and the saddle-node bifurcation takes place. However, due to the discreteness
of $n$, the first nonnegative value of $\hat\mu$ depends on $\hat\mu_s$ and
$\delta\mu$ (see Fig.~8). Now consider two values of $\delta\mu$, one
$\delta\mu_m$ satisfying $\hat\mu_s+m\,\delta\mu_m=0$, and another
$\delta\mu_{m+1}$ satisfying $\hat\mu_s+(m+1)\,\delta\mu_{m+1}=0$. Because
$\delta\mu_m$ and $\delta\mu_{m+1}$ are very close (for large $m$) {\it and}
both lead $\hat\mu(n)$ to pass through $\hat\mu=\hat\mu_*=0$ 
(one at time $n=m$, and
the other at time $n=m+1$), it is reasonable to assume that their orbits for
$\hat\mu_s/\delta\mu<n<n'$ are similar (except for a time shift
$n\rightarrow n+1$); i.e., they go to the same attractor. Thus, the period of
$1/\delta\mu$ is approximately $\Delta\,(1/\delta\mu)=1/\delta\mu_{m+1}-
1/\delta\mu_m=-\hat\mu_s^{-1}$.\\

We now consider the intervals of $1/\delta\mu$
between the centers of consecutive wide white bands in Fig.~6(b).
Figure 9 graphs the calculated fractal dimension $D'$ of the boundary between
white bands in these
consecutive intervals versus their center value of $1/\delta\mu$. From Fig.~9,
we see that as $1/\delta\mu$ increases, the graph of the
fractal dimension $D'$ does not converge to a definite value, but displays
further structure. Nevertheless, numerics show that as $1/\delta\mu$ becomes
large (i.e., in the range of $6.5\times 10^5$), $D'$ varies around the value
0.952. This is consistent with the numerics presented in Fig.~4(b) which
graphs the dimension of the fractal basin boundary for the time-independent
map $f_\mu$, at fixed values of the parameter $\mu$ where $\mu>\mu_*$. Thus, 
for large $1/\delta\mu$, $D'$ provides an estimate of the dimension of the 
fractal basin boundary in the absence of sweeping at $\mu>\mu_*$.\\

We now discuss a possible experimental application of our analysis.
The conceptually most straightforward method of measuring a fractal basin 
boundary would be to repeat many experiments each with precisely chosen initial
conditions. By determining the final attractor corresponding to each initial
condition, basins of attraction could conceivably be mapped out \cite{Virgin}. 
However, it is commonly the case that accurate control of initial conditions 
is not feasible for experiments. Thus, the application of this direct method 
is limited, and,
as a consequence, fractal basin boundaries have received little experimental
study, in spite of their fundamental importance. If a saddle-node
bifurcation occurs on the fractal basin boundary, an experiment can be 
arranged to take advantage of this. In this case, the purpose of the
experiment would be to measure the dimension $D'$ as an estimate of the fractal
dimension of the basin boundary $D$. The measurements would determine the 
final attractor of orbits starting at the attractor
to be destroyed by the saddle-node bifurcation, and swept through the
saddle-node bifurcation at different velocities (i.e., the
experimental data corresponding to the numerics in Fig.~6). This does not
require precise control of the initial conditions of the orbits. It is
sufficient for the initial condition to be 
in the basin of the attractor to be destroyed by the saddle-node
bifurcation; after enough time, the orbit will be as close to the
attractor as the noise level allows. Then, the orbit may be swept through
the saddle-node bifurcation. The final states of the orbits are
attractors; in their final states, orbits are robust to noise and to
measurement perturbations. The only parameters which require
rigorous control are the sweeping velocity (i.e., $\delta\mu$) and the initial 
value of the parameter
to be swept (i.e., $\mu_s$); precise knowledge of the parameter value where
the saddle-node bifurcation takes place (i.e., $\mu_*$) is not needed. 
[It is also required that the noise
level be sufficiently low (see Sec.~\ref{sec:noise}).]

\subsection{Capture Time}
\label{sec:capture}

A question of interest is how much time it takes for a swept orbit
to reach the final attracting state. Namely, we ask how many iterations with
$\mu>\mu_*$ are needed for the orbit to reach a neighborhood of the attractor
having the green basin. Due to slow sweeping, the location of the attractor
changes slightly on every iterate. If $x_\mu$ is a fixed point attractor of 
$f_\mu$ (with $\mu$ constant),
then a small change $\delta\mu$ in the parameter $\mu$, yields a change in
the position of the fixed point attractor,
\begin{eqnarray}
(x_{\mu+\delta\mu}-x_\mu)\equiv\delta x=\delta\mu\,
\frac{\frac{\partial f}{\partial \mu}(x_\mu;\mu)}
{1-\frac{\partial f_\mu}{\partial x}(x_\mu)}.\nonumber
\end{eqnarray}
We consider the swept orbit to have reached its final attractor if
consecutive iterates differ by about $\delta x$ (which is proportional to
$\delta\mu$). For numerical purposes, we consider that the orbit
has reached its final state if $|x_{n+1}-x_n|<10\,\delta\mu$. In our
numerical experiments, this condition is satisfied
by every orbit before $\mu$ reaches its final value $\mu_f$.
We refer to the number of iterations with $\mu>\mu_*$ needed to reach the
final state as the {\it capture time} of the corresponding orbit.
Figure 10 plots the capture time by the attractor $G_{\mu_f}$
[having the green basin in Fig.~2] versus $1/\delta\mu$ for a range
corresponding to one period of the structure in Fig.~6(b). No points are
plotted for values of $\delta\mu$ for which the orbit reaches the attractor 
$R_{\mu_f}$. The capture time graph has fractal features, since for
many values of $\delta\mu$ the orbit gets close to the fractal boundary
between $R[\mu]$ and $G[\mu]$. Using the fact that the final destination of the
orbit versus $1/\delta\mu$ is asymptotically periodic [see Fig.~6(b)], we can
 provide a further description of the capture time graph. We consider the
 series of the largest intervals of $1/\delta\mu$ for which the orbit reaches
 the attractor $G_{\mu_f}$ [see Fig.~6(b); we refer to the wide white band 
around
$1/\delta\mu=2400$ and the similar ones which are (asymptotically) separated
by an integer number of periods]. Orbits swept with $\delta\mu$ at the
centers of these intervals spend only a small number of iterations close to
the common fractal boundary of $R[\mu]$ and $G[\mu]$. Thus, the capture time
of such similar orbits does not depend on the structure of the fractal basin
boundary. We use Eq.~\eqref{eq:Ric} as an approximate description of these
orbits. A swept orbit reaches its final attracting state as $\hat x(n)$ becomes
large. Then, the orbit is rapidly trapped in the neighborhood of one of the
swept attractors of $f_\mu$. Thus, we equate the argument of the Airy
function in the denominator to its first root [see \eqref{eq:Ric}], solve for 
$n$, and substract $-\hat\mu_s/\delta\mu$ (the time for $\hat\mu$ to reach the
bifurcation value).
This yields the following approximate formula for the capture time
\begin{eqnarray}
\label{eq:capture}
n_C\approx |\tilde\xi|(a\,\delta\mu)^{-1/3},
\end{eqnarray}
where $\tilde\xi=-2.3381...$ is the largest root of the Airy function $Ai$.
Thus, we predict that for small $\delta\mu$, a log-log plot of the capture
time of the selected orbits versus $\delta\mu$ is a straight line with slope
 -1/3. Figure 11 shows the corresponding numerical results. The best fitting
line (not shown) has slope -0.31, in agreement with our prediction \cite{FN6}.

\subsection{Sweeping Through an Indeterminate Saddle-Node Bifurcation in the
Presence of Noise}
\label{sec:noise}
We now consider the addition of noise. Thus, we change our swept dynamical 
system to
\begin{eqnarray}
\label{eq:2Dmap_noise}
x_{n+1}&=&f_{\mu_n}(x_n)+A\,\epsilon_n,\\
\mu_{n}&=&\mu_s+n\,\delta\mu\nonumber,
\end{eqnarray}
where $\epsilon_n$ is random with uniform probability density in the interval
$[-1,1]$, and $A$ is a parameter which we call the noise amplitude. See 
Fig.~6(a) which shows the numerical results of the final destination of the 
orbits
versus $\delta\mu$ in the case $A=0$. The graph exhibits fractal features of
structure at arbitrarily small scales. The addition of small
noise is expected to alter this structure, switching the final destination of 
orbits. In this
case, it is appropriate to study the probability of orbits reaching one of the
final destinations. For every $A$, we compute the final attractor of a large
number of orbits having identical initial condition and parameters, but with
different realizations of the noise. We estimate the probability that an orbit
reaches a certain attractor by the fraction of such orbits that have reached 
the specified attractor in our numerical simulation. Figure 12 graphs the
probability that an orbit reaches
the attractor $G_{\mu_f}$ versus the noise amplitude $A$. We
present five graphs corresponding to five different values of $\delta\mu$
equally spaced in a range of $10^{-7}$ centered at $10^{-5}$
(i.e., $\delta\mu=10^{-5}$, $10^{-5}\pm 2.5\times 10^{-8}$ and
$10^{-5}\pm 5\times 10^{-8}$ ). We notice that
the probability graphs have different shapes, but a common
horizontal asymptote in the limit of large noise. The value of the horizontal
asymptote, approximately equal to 0.5, is related to the relative
measure of the corresponding basin.

As in the previous subsection, we take advantage of the asymptotically
periodic structure of the noiseless final destination graph versus 
$1/\delta\mu$
[see Fig.~6(b)]. We consider centers of the largest
intervals of $1/\delta\mu$ for which an orbit reaches the middle attractor in
the absence of noise.  We chose five such values of $\delta\mu$, spread over
two decades, where the ratio of consecutive values is approximately 3. Figure
 13(a) graphs the probability that an orbit reaches the middle fixed point
 attractor versus the noise amplitude $A$, for the five selected values of
$\delta\mu$. We notice that all the curves have qualitatively similar shape.
For a range from zero to small $A$, the probability is 1, and as $A$ 
increases,
the probability decreases to a horizontal asymptote.
The rightmost curve in the family corresponds to the largest value of
$\delta\mu$ ($\delta\mu\approx 3.445974\times 10^{-5}$), and the leftmost
curve corresponds to the smallest value of $\delta\mu$
($\delta\mu\approx 4.243522\times 10^{-7}$). Figure 13(b) shows the
same family of curves as in Fig.~13(a), but with the horizontal scale changed
from $A$ to $A/(\delta\mu)^{5/6}$. All data collapse to a single curve,
indicating that the probability that a swept orbit reaches the 
attractor $G_{\mu_f}$ depends only on the reduced variable 
$A/(\delta\mu)^{5/6}$. Later, we provide a theoretical argument for this 
scaling.

In order to gain some understanding of this result, we follow the idea of
Sec.~\ref{sec:sweep}, and use the canonical form $\hat f_{\hat \mu}$ to 
propose a simplified setup of our problem. We modify 
\eqref{eq:2Dmap} by the addition of a noise term $A\,\epsilon_n$ in the
right hand side of the first equation of \eqref{eq:2Dmap}. We are interested
in the probability that a swept orbit has at least one iterate, $\hat x_n$, in
a specified fixed interval far from the vicinity of the saddle-node 
bifurcation. More
precisely, we analyze how this probability changes versus $A$ and $\delta\mu$.
 Depending on the choice of interval and the choice of $\delta\mu$, 
the probability versus $A$ graph (not
 shown) has various shapes. For numerical purposes, we choose our fixed
 interval to be the same as that of Sec.~\ref{sec:sweep}, $100\leq\hat x\leq
250$. We then select values of $\delta\mu$ for which a noiseless swept orbit,
starting at $\hat x_s=-\sqrt{-\hat\mu_s/a}$, reaches exactly the center of
our fixed interval. The inverse of these values of $\delta\mu$ are centers of
intervals where the final state of the swept orbits is 0 [see Fig.~6(c)]. 
We consider five such values of $\delta\mu$, where the ratio of consecutive 
values is approximately 3. Figure 14(a) shows the probability that a swept 
orbit has an iterate in our fixed interval versus 
the noise amplitude for the selected values of $\delta\mu$.
Figure 14(a) shares the qualitative characteristics of
Fig.~13(a), with the only noticeable difference that the value of the
horizontal asymptote is now approximately 0.1. Figure 14(b) shows the same 
family of curves as in Fig.~14(a), where the horizontal scale has been changed
 from $A$ to $A/(\delta\mu)^{5/6}$. As for Fig.~12(b), this achieves good
 collapse of the family of curves.\\

We now present a theoretical argument for why the probability of reaching an 
attractor depends on $\delta\mu$ and $A$ only through the scaled variable
$A/(\delta\mu)^{5/6}$ when $\delta\mu$ and $A$ are small. From our results
in Figs.~14, we know that the scaling we wish to demonstrate should be
obtainable by use of the canonical form $\hat f_{\hat \mu}$.
Accordingly, we 
again use the differential equation approximation \eqref{eq:Ricatti}, but with
a noise term added,
\begin{eqnarray}
\label{eq:Ricatti_noise}
\frac{d\hat x}{d n}=n\,\delta\mu+a\hat x^2+A\hat\epsilon(n),
\end{eqnarray}
where $\hat\epsilon(n)$ is white noise,
\begin{eqnarray}
\langle \hat\epsilon(n)\rangle=0,\quad
\langle\hat\epsilon( n+ n')\hat \epsilon(n)\rangle=\delta( n'),\nonumber
\end{eqnarray}
and we have redefined the origin of the time variable $n$ so that the 
parameter $\hat\mu$ sweeps through zero at $n=0$ (i.e., we replaced
$n$ by $n-|\hat\mu_s|/\delta\mu$). Because we are only concerned
with scaling, and not with the exact solution of \eqref{eq:Ricatti_noise}, a 
fairly crude analysis will be sufficient.

First we consider the solution of \eqref{eq:Ricatti_noise} with the noise term
omitted, and the initial condition [see \eqref{eq:Ric_}]
\begin{eqnarray}
\hat x(0)=(-0.7290...)\left(\delta\mu/a^2\right)^{1/3}.\nonumber
\end{eqnarray}
We define a characteristic point of the orbit, 
$\hat x_{\rm nl}(n_{\rm nl})$, where 
$a\hat x_{\rm nl}^2\approx n_{\rm nl}\,\delta\mu$. For $n<n_{\rm nl}$,
$n\,\delta\mu\leq d\hat x/dn<2n\,\delta\mu$, and we can approximate the 
noiseless orbit as 
\begin{eqnarray}
\label{eq:xa}
\hat x(n)\approx \hat x(0)+\alpha(n)(n^2\,\delta\mu),
\end{eqnarray}
where $\alpha(n)$ is a slowly varying function of $n$ of order 1 
($1/2\leq \alpha (n)<1$ for $n<n_{\rm nl}$).
Setting $a\hat x^2\approx n\,\delta\mu$, we find that $n_{\rm nl}$ is given 
by 
\begin{eqnarray}
\label{eq:nl}
n_{\rm nl}\sim (a\,\delta\mu)^{-1/3},
\end{eqnarray}
corresponding to [c.f., Eq.~\eqref{eq:xa}] 
\begin{eqnarray}
\hat x_{\rm nl}\sim (\delta\mu/a)^{1/3}.\nonumber
\end{eqnarray}
For $n>n_{\rm nl}$ (i.e., $\hat x(n)>\hat x_{\rm nl}$), 
Eq.~\eqref{eq:Ricatti_noise} can be approximated as 
$d\hat x/dn\approx a\hat x^2$. 
Starting at $\hat x(n)\sim\hat x_{\rm nl}$, integration of this equation leads
to explosive growth of $\hat x$ to infinity in a time of order 
$(a\,\delta\mu)^{-1/3}$, which is of the same order as $n_{\rm nl}$. Thus,
the relevant time scale is $(a\,\delta\mu)^{-1/3}$ [this agrees with 
Eq.~\eqref{eq:capture} in Sec.~\ref{sec:capture}].

Now consider the action of noise. For $n<n_{\rm nl}$, we neglect the nonlinear
term $a\hat x^2$, so that \eqref{eq:Ricatti_noise} becomes 
$d\hat x/dn=n\,\delta\mu+A\hat\epsilon(n)$. The solution of this equation is
the linear superposition of the solutions of $d\hat x_a/dn=n\,\delta\mu$ and 
$d\hat x_b/dn=A\hat\epsilon(n)$, or $\hat x(n)=\hat x_a(n)+\hat x_b(n)$; 
$\hat x_a(n)$ is given by $\hat x_a(n)=\hat x(0)+n^2\delta\mu/2$, and 
$\hat x_b(n)$ is a random
walk. Thus, for $n<n_{\rm nl}$, there is diffusive spreading of 
the probability density of $\hat x$, 
\begin{eqnarray}
\label{eq:diff}
\Delta_{\rm diff}(n)
\equiv\sqrt{\langle\hat x_b^2(n)\rangle}\sim n^{1/2}A. 
\end{eqnarray}
This 
diffusive spreading can blur out the structure in Fig.~6. How large does the
noise amplitude $A$ have to be to do this? We can estimate $A$ by noting that
the periodic structure in Figs.~6(b,c) results from orbits that take different
integer times to reach $\hat x\sim \hat x_{\rm nl}$. Thus, for 
$n\approx n_{\rm nl}$ we define a scale $\Delta_{\rm nl}$ in $\hat x$ 
corresponding to the periodicity in $1/\delta\mu$ by [c.f., Eq.~\eqref{eq:xa}]
\begin{eqnarray}
\hat x_{\rm nl}\pm\Delta_{\rm nl}\approx
\hat x(0)+(n_{\rm nl}\pm 1)^2\delta\mu\nonumber
\end{eqnarray}
which yields 
\begin{eqnarray}
\label{eq:delxnl}
\Delta_{\rm nl}\sim n_{\rm nl}\delta\mu.
\end{eqnarray}
If by the time $n\approx n_{\rm nl}$, the diffusive spread of the probability
density of $\hat x$ becomes as large as $\Delta_{\rm nl}$, then the noise
starts to wash out the periodic variations with $1/\delta\mu$. Setting
$\Delta_{\rm diff}(n_{\rm nl})$ from \eqref{eq:diff} to be of the order of
$\Delta_{\rm nl}$ from \eqref{eq:delxnl}, we obtain 
$n_{\rm nl}^{1/2}A\sim n_{\rm nl}\delta\mu$, which with \eqref{eq:nl} yields
\begin{eqnarray}
A\sim (\delta\mu)^{5/6}.
\end{eqnarray}
Thus, we expect a collapse of the two parameter $(A,\delta\mu)$ data in 
Fig.~14(a) by means of a rescaling of $A$ by $\delta\mu$ raised to an 
exponent 5/6 [i.e., $A/(\delta\mu)^{5/6}$].

\section{Scaling of Indeterminate Saddle-Node Bifurcations for a Periodically
Forced Second Order Ordinary Differential Equation}
\label{sec:Duff}

In this section we demonstrate the scaling properties of sweeping through
an indeterminate saddle-node bifurcation in the case of the periodically
forced Duffing oscillator \cite{Breban},
\begin{eqnarray}
\label{eq:Duff}
\ddot x-0.15\,\dot x -x +x^3=\mu\cos t.
\end{eqnarray}
The unforced Duffing system (i.e., $\mu=0$) is an example of an
oscillator in a double well potential. It has two coexisting fixed point
attractors corresponding to the two minima of the potential energy.
For small $\mu$, the forced Duffing oscillator has two attracting
periodic orbits with the period of the forcing (i.e., $2\pi$), one in each 
well of the potential. At
$\mu=\mu_*\approx 0.2446$, a new attracting periodic orbit of period $6\pi$
arises through a saddle-node bifurcation. In Ref.~\cite{Aguirre}, it is argued
 numerically that for a certain range of $\mu>\mu_*$ the basin of
attraction of the $6\pi$ periodic orbit and the basins of attraction of the 
$2\pi$
periodic orbits have the Wada property. Thus, as $\mu$ decreases through
the critical value $\mu_*$, the period $6\pi$ attractor is destroyed
via a saddle-node bifurcation on the fractal boundary of the basins of the
other two attractors. This is an example of an indeterminate saddle-node
bifurcation of the Duffing system which we study by considering the 
two-dimensional map in the $(\dot x,x)$ plane resulting from a 
Poincar\'e section at constant phase of the forcing signal. We consider orbits
starting in the vicinity of the period three fixed point attractor, and, as we 
integrate the Duffing system, we decrease $\mu$ from $\mu_s>\mu_*$ to 
$\mu_f<\mu_*$ at a small rate of $\delta\mu$ per one period of the forcing
signal. As $\mu$ approaches $\mu_*$, (with $\mu>\mu_*$,) the period three 
fixed point attractor of the unswept Duffing system approaches its basin 
boundary, and the slowly swept orbit closely follows its location.
For $\mu-\mu_* <0$ small, the orbit will approximately follow the 
one-dimensional unstable manifold of the $\mu=\mu_*$ period three saddle-node 
pair. Thus, we can describe the sweeping 
through the indeterminate bifurcation of the Duffing oscillator by the theory
we developed for one dimensional discrete maps. Figure 15 shows the final 
destination graph of a swept orbit initially situated in the vicinity of the
period three fixed point of the Poincar\'e map. The final attracting state is 
represented as a 1 if situated in the potential well where $x>0$, and is 
represented as a 0
if situated in the potential well where $x<0$. As expected, the structure in
Fig.~15 appears asymptotically periodic if graphed versus $1/\delta\mu.$
In addition to slowly sweeping the Duffing system,
consider an additive noise term $A\,\epsilon(t)$ in the right
hand side of \eqref{eq:Duff}, where on every time step $\epsilon(t)$ is
chosen randomly in $[-1,1]$, and the time step used is 
$\Delta t=2\pi/500$. Figure 16(a) shows the dependence of the 
probability of approaching the attractor represented as a 1 
versus the noise amplitude $A$
for three specially selected values of $\delta\mu$ (centers of white bands in 
the structure of Fig.~15 where the swept orbit reaches the attracting state 
represented by 1) spread over one decade. Figure 16(b) shows collapse of the 
data in Fig.~16(a) to a single curve when the noise amplitude $A$ is rescaled 
by $(\delta\mu)^{5/6}$, as predicted by our previous one-dimensional analysis
(Sec.~\ref{sec:noise}).
Thus, we believe that the scaling properties of the indeterminate saddle-node 
bifurcation we found in one-dimensional discrete maps are also shared by 
higher dimensional flows.

\section{Indeterminacy in How an Attractor is Approached}
\label{sec:model2}

In this section we consider the case of a one dimensional map $f_\mu$ having
two attractors A and B, one of which (i.e., A) exists for all 
$\mu\in [\mu_s,\mu_f]$. The other
(i.e., B) is a node which is destroyed by a saddle-node bifurcation on the 
boundary between the basins of A and B, as $\mu$ increases through $\mu_*$
($\mu_*\in [\mu_s,\mu_f]$).
When an orbit is initially on B, and $\mu$ is slowly increased through 
$\mu_*$,
the orbit will always go to A (which is the only attractor for $\mu>\mu_*$).
However, it is possible to distinguish between two (or more) different ways
of approaching A. [In particular, we are interested in ways of approach that 
can be distinguished in a coordinate-free 
(i.e., invariant) manner.] As we show in this section, the way in
which A is approached can be indeterminate. In this case, the indeterminacy 
is connected with the existence of an invariant nonattracting Cantor set 
embedded in the basin of A for $\mu>\mu_*$.

As an illustration, we construct the following model
\begin{eqnarray}
\label{eq:model2}
f_\mu(x)=-\mu +x-3x^2-x^4+3.6x^6-x^8.
\end{eqnarray}
Calculations show that $f_\mu$ satisfies all the requirements of
the saddle-node bifurcation theorem for
undergoing a backward saddle-node bifurcation at $x_*=0$ and $\mu_*=0$.
Figure 17(a) shows the graph of $f_\mu$ versus $x$ at $\mu =\mu_*$. Figure
17(b) shows how the basin structure of the map $f_\mu$ varies with the
parameter $\mu$. For positive values of $\mu$, $f_\mu$ has only one
attractor which is at minus infinity. The basin of this
attractor is the whole real axis. As $\mu$ decreases through $\mu_*=0$, a new
fixed point attractor is created at $x_*=0$. The basin of attraction of this 
fixed point has infinitely many disjoint intervals displaying fractal features
 [indicated in black in
 Fig.~17(b)]. This is similar to the blue basin $B[\mu]$ of the attractor 
$B_\mu$ of the previous one-dimensional model (see Sec.~\ref{model1}).

The blue region in Fig.~18(a) is the basin of attraction of the stable fixed
point destroyed as $\mu$ increases through $\mu_*$. For every value
of $\mu$ we consider, the map $f_\mu$ has invariant Cantor sets. The
trajectories
of points which are located on an invariant Cantor set, do not diverge to
infinity. One way to display such Cantor sets, is to select uniquely defined
intervals whose end points are on the Cantor set. For example, Fig.~18(a)
shows green and red regions. For every fixed parameter value $\mu$, the
collection of points that are boundary points of the red and green regions,
constitutes an invariant Cantor set. In order to describe these green and red
regions, we introduce the following notations. For each parameter value $\mu$,
let $p_\mu$ be the leftmost fixed point of $f_\mu$ [see Fig.~17(a)]. For
every $x_0<p_\mu$, the sequence of iterates $\{x_n=f_\mu^{[n]}(x_0)\}$ is
decreasing and diverges
to minus infinity. For each value of $\mu$, let $q_\mu$ be the fixed
point of $f_\mu$ to the right of $x = 0$ at which $\frac{\partial f_\mu}
{\partial x}(q_\mu)>1$. A point $(x;\mu)$
is colored green if its trajectory diverges to minus infinity and it passes
through the
interval $(q_\mu ,\infty)$, and it is colored red if its trajectory diverges
to minus infinity and it does not pass through the interval $(q_\mu,\infty)$.
Denote the collection of points $(x;\mu)$ that are colored green by $G[\mu]$,
and the collection of points $(x;\mu)$ that are colored red by $R[\mu]$.
Using the methods and techniques of \cite{N}, it can be shown that the
collection of points $(x;\mu)$ which are common boundary points of $G[\mu]$
and $R[\mu]$ is a Cantor set $C[\mu]$ \cite{FN1}. In particular, the results
of \cite{N} imply that for $\mu=\mu_*=0$ the point $x_*=0$
belongs to the invariant Cantor set $C[\mu_*]$.

Figure 18(b) is a zoom of Figure 18(a) in the
region of the saddle-node bifurcation. For values of $\mu>\mu_*$, in the
vicinity of $(x_*;\mu_*)$, one notices a fractal alternation of red and green
stripes. The green and red stripe structure in Fig.~18(b) shares
 qualitative properties with the structure in Fig.~2(b). All
the analysis in Sec.~\ref{sec2} can be adapted straightfowardly to fit
this situation.

Figure 19 shows how the chaotic saddle of the map $f_\mu$ varies with $\mu$.
The chaotic saddle is generated numerically using the PIM-triple method. For
an explanation of this method see Nusse and Yorke
\cite{NY2}. Using arguments similar to those in Sec.~\ref{sec:scdim}, we
predict that changing the horizontal axis of Fig.~19 from $\mu$ to
$(\mu-\mu_*)^{-1/2}$ makes the chaotic saddle asymptotically periodic.
Numerical results confirming this are presented in Fig.~20(a).
For $f_\mu$ given by \eqref{eq:model2}, we were able to find a parameter 
value 
$\mu_{**}= 0.23495384$ where changing the horizontal axis of Fig.~19
from $\mu$ to $(\mu-\mu_{**})^{-1/2}$ [see Fig.~20(b)] apparently makes the
chaotic saddle asymptotically periodic [with a different period than that of 
Fig.~20(a)]. As in the case discussed in Sec.~\ref{sec2}, past the
saddle-node bifurcation of $f_\mu$ at $\mu_*$,
infinitely many other saddle-node bifurcations of periodic orbits 
take place on the invariant Cantor set $C[\mu]$. We believe that $\mu_{**}$
is an approximate value of $\mu$ where such a saddle-node of a periodic orbit 
takes place.

\section{Discussion and Conclusions}

In this paper, we have investigated scaling properties of saddle-node 
bifurcations that occur on fractal basin boundaries.
Such situations are known to be indeterminate in the sense that it is 
difficult to predict the eventual fate of an orbit that tracks the 
pre-bifurcation node attractor as the system parameter is swept through the 
bifurcation.
We have first analyzed the case of one-dimensional discrete maps. 
Using the normal form of the 
saddle-node bifurcation and general properties of fractal basin boundaries,
we established the following universal (i.e., model independent) scaling
results
\begin{itemize}
\item scaling of the fractal
basin boundary of the static (i.e., unswept) system near the saddle-node 
bifurcation,
\item the scaling dependence of the orbit's final destination with the 
inverse of the sweeping rate,
\item the dependence of the time it takes for an attractor to 
capture a swept orbit with the -1/3 power of the sweeping rate,
\item scaling of the effect of noise on the  
final attractor capture probability with the 5/6 power of the sweeping rate.
\end{itemize}
All these results were demonstrated numerically for a one-dimensional map 
example. Following our one-dimensional investigations,
we have explained and demonstrated numerically that these new results 
also apply to two-dimensional maps. Our numerical example was a 
two-dimensional
map that results from a Poincar\'e section of the forced Duffing oscillator.
In the last section of the paper, we have discussed how the new results 
listed above
apply to the case where a saddle-node bifurcation occurs on an invariant
Cantor set which is embedded in a basin of attraction, and we have supported 
our discussion by numerics.

This work was supported by ONR (Physics), NSF (PHYS 0098632), NSF (0104087), 
and by the W.M. Keck Foundation.

\newpage
\begin{figure}
\begin{center}
\end{center}
\caption{Construction of
the function $f_\mu(x)$ starting with (a) the third iterate of the logistic 
map, $g(x)=r\,x(1-x)$, with $r=3.832$, and adding a perturbation (b)
$\mu \sin(3\pi x)$ ($\mu =5.4\times 10^{-3}$).}
\end{figure}

\begin{figure}
\begin{center}
\end{center}
\caption{
 (a) Basin structure of the map $f_\mu$ versus the
parameter $\mu $ on the horizontal axis ($0\leq\mu\leq 5.4\times 10^{-3}$
and $0\leq x\leq 1$). The attractor having the blue basin is
destroyed at $\mu \approx 2.79\times 10^{-3}$. (b) Detail of the region
shown as the white rectangle in Fig.~2(a), $2.75\times 10^{-3}\leq\mu
\leq 3.55\times 10^{-3}$ and $0.145\leq x\leq 0.163$.}
\end{figure}

\begin{figure}
\begin{center}
\end{center}
\caption{Fractal dimension of the basin boundary versus $\mu$. Notice the
continuous variation for $\mu<\mu_*$ and the discontinuous jump at $\mu_*$, the
parameter value at which the saddle-node bifurcation on the fractal basin
boundary takes place.}
\end{figure}

\begin{figure}
\begin{center}
\end{center}
\caption{
(a) Detail of Figure 2(b), with the horizontal axis changed from $\mu$ to
$(\mu-\mu_*)^{-1/2}$ for $\mu>\mu_*$;
$2.75\times 10^{-3}\leq\mu\leq 3.55\times10^{-3}$
 and $0.145\leq x\leq 0.163$
The green stripes from Figure 2(b) are colored black and the red stripes are
colored white. The approximate position of the point $x_*$ where the
saddle-node bifurcation takes place is shown. $x_c$ indicates the nearest
critical point. (b) Detail of Figure 3, displaying how the box dimension $D$
of the fractal basin boundary varies with $1/(\mu_*-\mu)^{1/2}$.
The horizontal axis of Figures 4(a) and 4(b) are identical.}
\end{figure}

\begin{figure}
\begin{center}
\end{center}
\caption{Qualitative graphs of the solution of Eq.~\eqref{trans},
$\hat{\mu}_n^{-1/2}(\hat{x}_0)$, for three consecutive values of $n$.
Note the horizontal asymptotes [$\hat\mu^{-1/2}=(n-1)a^{1/2}\pi$,
$n\,a^{1/2}\pi$, and $(n+1)a^{1/2}\pi$], the vertical asymptotes
[$\hat x_s=(a(n-1))^{-1}$, $(an)^{-1}$, and $(a(n+1))^{-1}$], both shown as
dashed lines, and the intersections of the solid curves with $\hat{x}_0=0$
 which are marked with black dots.}
\end{figure}

\begin{figure}
\caption{(a) Final attracting state of swept orbits versus $\delta\mu$.
We have chosen $\mu_s=\hat\mu_s+\mu_*=0$,
 and $\mu_f=4.5\times 10^{-3}$. The attractor $R_{\mu_f}$ is represented
  by 1 and the attractor  $G_{\mu_f}$ is represented by 0.
(b) Detail of Fig.~6(b) with the horizontal scale changed from $\delta\mu$ to
$1/\delta\mu$. The structure of white and black bands becomes asymptotically
periodic. (c) Final state of orbits for the system $\hat f_{\hat \mu}$ versus
 $1/\delta\mu$. The final state of an orbit is defined to be 0 if there
 exists $n$ such that $100<\hat x_n<250$, and is defined to be 1, otherwise.
 We have chosen $\hat\mu_s=-\mu_*$, so that Figs.~6(b,c) have the same
 asymptotic periodicity.}
\end{figure}

\begin{figure}
\begin{center}
\end{center}
\caption{Numerical results for the inverse of the limit period in $1/\delta\mu$
versus $\mu_s$. The fit line is
$[\Delta\,(1/\delta\mu)]^{-1}=-0.9986\mu_s+0.0028$ and indicates good
agreement with the theoretical explanation presented in text.}
\end{figure}

\begin{figure}
\begin{center}
\end{center}
\caption{Graphs of $\hat f_{\hat \mu}(\hat x)$ at different values of the
parameter $\hat \mu$. The black dots indicate the stable fixed points of
$\hat f_{\hat \mu}$ for different values of $\hat \mu$.}
\end{figure}

\begin{figure}
\begin{center}
\end{center}
\caption{The calculated fractal dimension $D'$ of the structure in the 
intervals between the centers of consecutive wide white bands in Fig.~6(b)
versus their center value of $1/\delta\mu$.}
\end{figure}

\begin{figure}
\begin{center}
\end{center}
\caption{Capture time by the fixed point attractor $G_{\mu_f}$
 versus $1/\delta\mu$. We have chosen $\mu_s=0$. The range of
$1/\delta\mu$ is approximatelly one period of the graph in Fig.~6(b), with
$\delta\mu\approx 10^{-8}$. The vertical axis
ranges between 250 and 650. No points are plotted for values of $\delta\mu$
for which the orbit reaches the fixed point attractor 
$R_{\mu_f}$.}
\end{figure}

\begin{figure}
\begin{center}
\end{center}
\caption{Capture time by the middle fixed point attractor of $f_{\mu}$
 versus $\delta\mu$ ($\mu_s=0$). The best fitting line
(not shown) has slope -0.31, in agreement with the
theory.}
\end{figure}

\begin{figure}
\begin{center}
\end{center}
\caption{Probability that one orbit reaches the middle fixed point attractor
of $f_{\mu}$ versus the noise amplitude $A$, for five different values of
$\delta\mu$ ($10^{-5}$, $10^{-5}\pm 2.5\times 10^{-8}$ and
$10^{-5}\pm 5\times 10^{-8}$). We have chosen $\mu_s=0$.}
\end{figure}

\begin{figure}
\begin{center}
\end{center}
\caption{Probability that an orbit reaches the middle fixed point attractor
of $f_{\mu}$, for five selected values of
$\delta\mu$ spread over two decades:
 (a) versus the noise amplitude $A$, and (b) versus $A/(\delta\mu)^{5/6}$,
 We have chosen $\mu_s=0$.
 From right to left, the $\delta\mu$ values corresponding to the curves are
 approximately: 3.445974$\times 10^{-5}$, 1.147767$\times 10^{-5}$,
 3.820744$\times 10^{-6}$,
 1.273160$\times 10^{-6}$ and 4.243522$\times 10^{-7}$.}
 \end{figure}

\begin{figure}
\begin{center}
\end{center}
\caption{Probability that an orbit of $\hat f_{\hat \mu}$ reaches a fixed
interval far from the saddle-node bifurcation (i.e., [100, 250]),
 for five values of $\delta\mu$ spread over two decades:
 (a) versus the noise amplitude $A$, and (b) versus $A/(\delta\mu)^{5/6}$.
We have chosen $\mu_s=0$.
 From right to left, the $\delta\mu$ values corresponding to the curves are
 approximately: 3.451540$\times 10^{-5}$, 1.149162$\times 10^{-5}$,
 3.829769$\times 10^{-6}$,
 1.276061$\times 10^{-6}$ and 4.253018$\times 10^{-7}$.}
\end{figure}

\begin{figure}
\begin{center}
\end{center}
\caption{Final attracting state of swept orbits of the Duffing oscilator
versus $1/\delta\mu$. The structure of white and black bands becomes
asymptotically periodic. We have chosen $\mu_s=0.253$,
 and $\mu_f=0.22$. The attractor in the potential well for $x>0$
is represented as a 1, and the attractor in the potential well for $x<0$ is 
represented as a 0.}
\end{figure}

\begin{figure}
\begin{center}
\end{center}
\caption{Probability the Duffing oscillator reaches
the attracting periodic orbit in the potential well at $x>0$
 for three values of $\delta\mu$ spread over one decades:
 (a) versus the noise amplitude $A$, and (b) versus $A/(\delta\mu)^{5/6}$.
We have chosen $\mu_s=0.253$.
 From right to left, the $\delta\mu$ values corresponding to the curves are
 approximately: 4.628716$\times 10^{-5}$,
 1.461574$\times 10^{-5}$ and 4.621737$\times 10^{-6}$.}
\end{figure}

\begin{figure}
\begin{center}
\end{center}
\caption{(a) Graph of $f_\mu(x)$
 versus x at the bifurcation parameter. (b) Basin structure of map $f_\mu(x)$
versus the parameter $\mu $ ($-0.3\leq\mu\leq 0.3$ and
$-2\leq x\leq 2$). The basin of attraction of the stable fixed point created
by the saddle-node bifurcation is black while the basin of attraction of
minus infinity is left white.}
\end{figure}

\clearpage
\begin{figure}
\begin{center}
\end{center}
\caption{(a) Basin structure of
 $f_\mu$ versus $\mu $ ($-0.3\leq\mu\leq 0.3$ and
$-2\leq x\leq 2$). We split the basin of attraction of minus infinity into two
components, one plotted as the green region and the other plotted as the red
region. The green region is the collection of all points that go to
minus infinity and have at least one iterate bigger that the unstable fixed
point $q_\mu$. The red set is the region of all the other points that go to
minus infinity. (b) Detail of Fig.~15(a) in the region shown as the white
rectangle, $-0.005\leq\mu\leq 0.015$ and $-0.09\leq x\leq 0.41$.}
\end{figure}

\clearpage
\begin{figure}
\begin{center}
\end{center}
\caption{The chaotic saddle of $f_\mu$ versus $\mu $
($-0.3\leq\mu\leq 0.3$ and $-2\leq x\leq 2$)
generated by the PIM-triple method.}
\end{figure}

\begin{figure}
\begin{center}
\end{center}
\caption{The chaotic saddle of the map $f_\mu$ in the vicinity of the
saddle-node bifurcation with the horizontal axis rescaled from $\mu $ to:
(a) $(\mu_*-\mu)^{-1/2}$. Notice that the chaotic saddle becomes
asymptotically periodic
($-0.008\leq x\leq 0.337,\, 10\leq (\mu_*-\mu)^{-1/2}\leq 
15)$.
(b) $(\mu_{**}-\mu)^{-1/2}$, where $\mu_{**}=0.23495384$. We believe that
 $\mu_{**}$ corresponds to the approximate value of the parameter $\mu$ where 
a saddle-node bifurcation of a periodic orbit of $f_\mu$ takes place on the
Cantor set $C[\mu]$. In this case, the
chaotic saddle also becomes asymptotically periodic 
($-0.162\leq x\leq 0.168,\, 9.97 < (\mu_{**}-\mu)^{-1/2} < 2010 $).}
\end{figure}


\begin{thebibliography}{6}
\bibitem{FN6} For numerical purposes, we consider that a swept orbit has 
reached
its final state if $|x_{n+1}-x_n|<10\,\delta\mu$. Changing the factor of 10
to other values ranging from 1 to 30, can slightly change the numerical result
for the slope (by as much as 3\%).
\bibitem{FN1} For every $\mu$ ($-0.3 < \mu < 0.3$), write
$p^*_\mu = \max \{x \in \mathbb{R}: f_\mu (x) = p_\mu \}$ and
$q^*_\mu = \max \{x \in \mathbb{R}: f_\mu (x) = q_\mu \}$. 
The interval $[q_\mu ,q^*_\mu]$ also contains a Cantor set. By coloring this
whole segment green, this information is lost. Therefore, the coloring scheme
should be adapted if one wants to have the whole invariant Cantor set
represented for every $\mu$. For example, if a trajectory that diverges to
minus infinity contains a point that is greater than $p^*_\mu$ then the
initial point is colored green, if a trajectory that diverges to minus
infinity contains a point that is greater than  $q^*_\mu$ but not greater than
$p^*_\mu$ then the initial point is colored yellow. A
point is colored red, if its trajectory diverges to minus infinity
and does not have a point that is greater than $q^*_\mu$. Then the collection
of boundary points (a point $x$ is a boundary point if every open
neighborhood of $x$ contains points of at least two different colors) is a
Cantor set that contains the Cantor set described above.
\end{thebibliography}
\end{document}